# Distributed Self Management for Distributed Security Systems

Michael Hilker

University of Luxembourg
Faculty of Science, Technology, and Communications
6, Rue Richard Coudenhove-Kalergi, L-1359 Luxembourg
`michael.hilker@uni.lu`

**Abstract.** Distributed system as e.g. artificial immune systems, complex adaptive systems, or multi-agent systems are widely used in Computer Science, e.g. for network security, optimisations, or simulations. In these systems, small entities move through the network and perform certain tasks. At some time, the entities move to another place and require therefore information where to move is most profitable. Common used systems do not provide any information or use a centralised approach where a center delegates the entities. This article discusses whether small information about the neighbours enhances the performance of the overall system or not. Therefore, two information-protocols are introduced and analysed. In addition, the protocols are implemented and tested using the artificial immune system SANA that protects a network against intrusions.

## 1 Introduction

Distributed systems are widely used in computer science in order to solve problems distributed on several nodes. Examples are simulation or optimisation problems, which run on several nodes in parallel or the network security where a network is secured against intrusion like e.g. viruses, worms, and trojans. These distributed systems are e.g. multi-agent systems [1], complex adaptive systems [2], and artificial immune systems [3,4,5]. These systems use small entities - e.g. agents or artificial cells - that move through the network, perform certain tasks for network security, and work mostly autonomously. These entities receive local information about the situation in the network and contain a limited knowledge about the overall goal - here secure a network against intrusions. An example is that an entity knows how to identify one certain intrusion and receives the status of the node where the entity resides [6]. The performance of the overall systems originates from the performance of all the small entities, which leads to waste of computational power and decreased performance of the overall system when the entities are not properly distributed over all nodes.

A key problem in the distributed systems is that the entities work autonomously and have to decide in some timestep whether they move to another node and if yes to which node - this problem is part of the self-management of a distributed system. This problem is significant because the distributed security system should secure the whole network and all nodes should have at least a certain number of entities. This article discusses whether exchanging a small piece of information about the nearby situation increases the performance of the security system or not - local information about a node with some additional global information about the nearby region of the network. Therefore, two information protocols describing how and which information is exchanged are introduced. Thereafter, the two approaches are implemented in an artificial immune system that uses autonomous artificial cells in order to secure a network against intrusions. Lastly, some discussions about the test results and the analysis conclude the paper.

The domain of the application is network security. It describes the research in order to secure a network against intrusions. Intrusions are e.g. viruses, worms, and trojans. There are several approaches in order to secure a network against intrusions, e.g. antivirus software, packet filters, and intrusion detection systems [7]. Most of these common used security systems lack from the local installation because then they only secure one network node and require lots of computational power to check the whole network traffic. Furthermore, the security solutions for a network consist of several systems, e.g. antivirus software in each node, packet filters in each router, and intrusion detection systems at the internet gateway and email server. However, these systems are not connected and do not work together. In order to protect a network against the upcoming intrusions, a more sophisticated approach is required where an example are the distributed systems. Thus, a distributed approach is chosen in the artificial immune system SANA [4] where artificial cells perform the required tasks and build up a distributed intrusion detection system. These cells - the entities - move through the network and require the information where to roam next. The next section introduces the existing approaches for sharing movement information in distributed systems.

## 2 Current situation

Currently, there are two different approaches in order to manage the movement of the entities in a distributed system:

1. The entities do not work fully autonomously. A centralised system, which knows the situation of the whole network, delegates the entities - mostly agents - to their destinations. Thus, the system is again dependent on a centralised system, which is also a single point of failure and is a contradiction to the idea of a distributed system.
2. When the entities work fully autonomously, in most systems they receive no information when and where to move. The movement is normally modelled using some probabilistic events with a distribution over all neighbour nodes.

Both systems have significant disadvantages. The centralised approach requires lots of bandwidth because the center needs information from all over the network and the center is a single point of failure. The distributed approach reduces the bandwidth significantly but it is possible that the entities move to inappropriate nodes where e.g., some nodes have too many entities and other nodes suffer from the low number of entities. Consequently, this article discusses how small pieces of information can help the entities to do the right decision.

## 3 Information Management and Protocols

For investigating research in the information management, the following situation in the network is assumed:

An artificial immune system [4] with lots of artificial cells - the entities - protects a network against intrusions [6]. Therefore, a type of artificial cell knows how to detect and remove exactly one intrusion where there are several instances of this type of artificial cells, cp. the immune cells in the human immune system [8]. The whole system does not contain any centralised system and the artificial cells decide autonomously when and where to move. The artificial immune system presents each packet to all artificial cells in this node and the cells decide whether the packet contains an intrusion or not. These cells should be distributed over all nodes so that each node contains approximately the same number of cells. In addition, the artificial immune system contains artificial cells that move through the network and check the nodes whether they are infected by an intrusion or not; e.g. in looking for backdoors or significant files and processes on the node. These artificial cells require the information where to go because the checks should be done uniformly in time distributed over all nodes; this means that each node is checked regularly but not too often. This problem is the same as the general problem of the entities moving to another node. The artificial cells are the entities and the infrastructure provides the required resources for enabling the information management.

For the first problem - the movement of artificial cells that check packets - the notification of insufficient security is introduced and for the second problem - the movement of artificial cells that check nodes - the trails of entities is introduced. In both approaches, the required resources must be limited. In the first approach, only one network packet is at most required for each connection in the network. The second approach requires only some storage in the network nodes that depends on the number of different types of artificial cells. Furthermore, the artificial cells receive some small pieces of information and, henceforth, do not increase that much in size and complexity.

### 3.1 Notification of insufficient Security

In this approach, the problem of entities that check packets is discussed [9]. In general, this approach can be used for all entities, which perform regularly certain tasks and which are required in lots of nodes of the network. The approach is motivated by the cytokines and hormones of the humen immune system regulating the immune response and its immune cells motivate this approach [8].

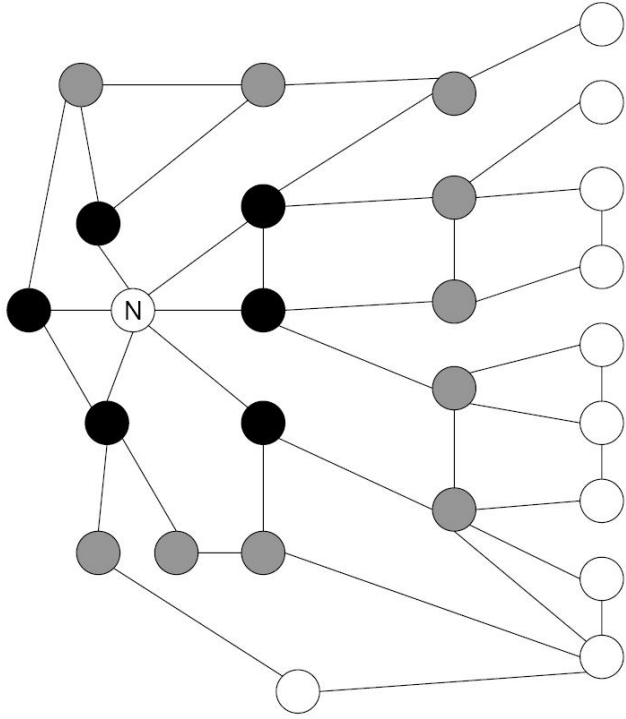

**Fig. 1.** Distribution of Notification: The node marked with $N$ has an insufficient security level and notifies this to its neighbours. The receiving time of this information is indicated using the colour of the nodes where the more black the earlier the information arrives. The nodes without a colour do not receive the information because they are too far away.

For detecting that a node $n$ has an insufficient number of entities, it has a parameter that represents the minimum number of required security $min\_sec_{node}$. Each entity returns on demand how many security it provides: $sec_{entity}$. The node $n$ asks every time step all entities residing in this node how many security each entity provides and sums this up: $sec_{time} = \sum\limits_{e \in entity\_node} sec_e$; a weighted sum for a more sophisticated calculation can be also used. If the security in this time step is below the required security $min\_sec_n > sec_{time}$, the node $n$ starts a notification process in order to allure nearby entities.

This notification works as follows:

1. Using local communication, the node informs the entities that are working in this node about the insufficient security level. These entities are then more affine to stay than to move. However, the entities are not forced to stay, they still decide on their own.

2. The node furthermore sends a network packet to all neighbours that indicate the low security level and the packet also describes the missing level.
3. The nodes receiving this packet forwards it to all entities working in these nodes. These entities then get more affine to move to this node than to stay or to move to another node. The change of the affinity is dependent on the level of lacking security that is stored in the packet.
4. With this workflow, each node maybe receives each time step several packets describing insufficient security levels of several nodes. Therefore, the node will forward only the packet with the highest missing security level.
5. In order to receive a global notification, each node forwards these notification packets to their neighbours. The value describing the missing value is therefore decreased by a decreasing function $f_d$ and if the value is still above a threshold, the packet is forwarded to all neighbours with the decreased value.
6. The notification about insufficient security in a node is not stored.

The distribution of the information that a node has insufficient security level is visualised in figure 1.

The entities - here artificial cells - react to this information as follows. Each entity has a function $f_m$ that describes the probability that it moves to another node. Another function $f_w$ describes where the entity moves next. These functions can have several parameters as the time the entity did not move, the network load, and lacking security level. Therefore, each time step a probabilistic event is done that succeeds with probability $f_m$ and if this event succeeds a second probabilistic event using $f_w$ describes the node where the entity will move. When the entity receives a notification that a node has an insufficient value, the function $f_m$ will be increased in order to increase the affinity to move and the second function $f_w$ is so changed that the entity will always move towards the node with insufficient security. The function $f_m$ is visualised in figure 2.

This approach is implemented in the artificial immune system SANA that is developed at the University of Luxembourg for the purpose of network security (section 6). Each node with low security level sends the notification each time step and the entities react to it. In the simulations, the functions are set as follows:

- $min\_sec_{node}$ is set to 20 where there are 60 different types of artificial cells in the artificial immune system
- $sec_{entity}$ is the number of known intrusions, mostly 1
- the decreasing function $f_d$ is set to $old\_value - 1$
- if a node has an insufficient security level the artificial cells in this node will not leave it
- the movement function $f_m$ is shown in figure 2
- the where to move function $f_w$ is only used if there is not any notification, otherwise the entities move towards this node

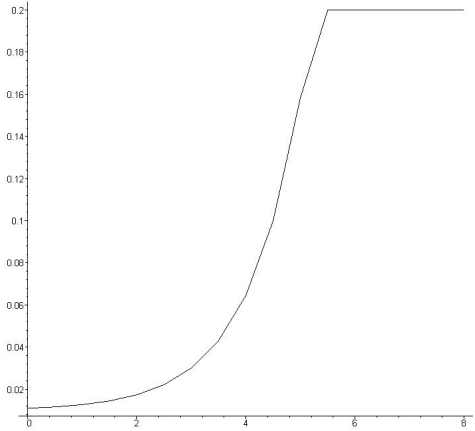

**Fig. 2.** Response function $f_m$ of an entity: $f_m(0)$ describe the normal affinity to move; $f_m(x)$ describes the moving probability when lacking security $x$ is received; maximum limits response to prevent overreaction.

### 3.2 Trails of Entities

This approach provides information to entities performing regular checks in nodes as identifying infected nodes. In general, the approach can be used for entities performing a tasks that must be done regularly but not too often in a node. Therefore, an approach is used that is motivated by artificial ant colonies [10].

Each node $n$ contains storage for each outgoing connection $(v, w)$. In this storage, for each type of entity $e$ is a value stored $v_{v,w,e}$ that describes when the last check took place. When an entity does a check in the node and moves to the next node, it increases this value $v_{v,w,e}$ and every time step the node decreases this value $v_{v,w,e}$ until it reaches zero. The entities in this node use this information in order to evaluate where to move next.

Assume the following situation: an entity is in node $n$ and finishes the check. The next step is to move to another node and the entity receives information when an entity of the same type moves to the neighbours, i.e. the values $v_{v,w,e}$ for each $(v, w)$ in $n$ for type $e$. Then, it uses this information for a probabilistic event where the $Pr$(move to neighbour $n_i$) depends on the value that describes when the node $n_i$ was lastly checked. In addition, the value is not a counting of time steps. For the increasing of the value is a function $f_i$ and for the decrease a function $f_d$ used. Then it is possible to when the last check did place at the neighbouring node as well as how often in the past the node was checked by this type of artificial cell. An example of $f_i$ is discussed in figure 3.

This approach is also implemented in the artificial immune system where the storage uses float-values. The increasing function $f_i$ is set to $c_1 + c_2 \cdot exp(old\_value)$ with $c_1$ and $c_2$ as a constant value, e.g. $c_1 = 10$ and $c_2 = \frac{1}{1000}$. The decreasing function $f_d$ is set to $old\_value - c_3$ where $c_3$ is a constant value, e.g. 2. Thus, the value increases exponentially and decreases linearly. The probabilistic event sets the probability for each node dependent on the required security in the neighbour nodes.

## 4 Testing Results

As already explained, the two techniques are implemented in an artificial immune system. The simulations are done using a network simulator implementing a packet-switching network. The network is a company-like network with about 500 nodes with different server, workstations, and services. An Internet gateway connects the network to the Internet and different intrusions are implemented, which are mostly viruses, worms, and trojans arriving over the network or directly by internal attacks. The results are discussed in the next two subsections:

### 4.1 Notification of insufficient Security

The notification of insufficient security works more than acceptable. When the parameters and functions are chosen properly, the system adapts quickly so that nodes with insufficient security receive artificial cells quickly. Furthermore, the system does not overreact which means that the artificial cells react too heavily and all cells move immediately to a node with insufficient security. Then, the alternating status is prevented: this means that all cells oscillate between two nodes because one of these two nodes has always an insufficient security level.

The required resources are limited. For each connection is exactly one network packet for each time step used. The required computational power is also limited because the processing of the information is quite fast. The performance increase of the artificial immune system is observed. The performance with notification of insufficient security levels is significant higher than in the version without enabled notification. For example in one scenario, about 80% of the intrusions are identifies without the usage of the notification system; this low value occurs because some nodes have too much and some nodes too few artificial cells. When the notification system is enabled, the system identifies about 98% of the intrusions because all nodes have a guaranteed security value. Compared to the centralised approach with an entity management server, the notification of insufficient security requires about 80% less bandwidth and the performance increases by 15% due to the quicker adaption to weak points in the network.

The reason for this is that the system quickly solves security problems and the attackers cannot use nodes with insufficient security in order to install intrusions. Additionally, the system distributes the artificial cells in a better way. If there is a node with too many artificial cells that requires too much computational power, there are also nodes with insufficient security and the artificial cells will solve this problem using the information. Furthermore, it is possible to adapt the system to the current situation using changes of the parameters. In addition, important nodes are secured with more entities, which is configured through changes in the notification system.

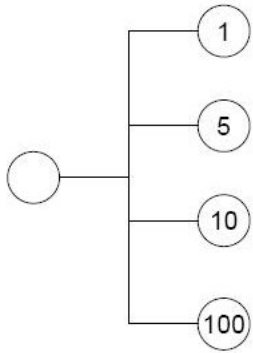

**Fig. 3.** Trails: an entity in the left node wants to move, reads therefore the trails, and receives the values indicating the trail. Thereafter, it makes a probabilistic event where it chooses uniformly a number from $[1, 289]$. If the number is in $[1, 100]$ it moves to the node with trail value 1. Else, if the number is in $[101, 197]$, it moves to the node with trail value 5. Else, if the number is in $[198, 288]$, it moves to the node with trail value 10. Else, if the number is 289, it moves to the node with trail value 100. This probabilistic event reflects the trail values of the neighbouring nodes.

### 4.2 Trails of Entities

The trails of entities work well. The system provides information to the entities where their tasks are required and the entities move there. The entities also save computational power because between two checks is a certain time period. In our simulations, without the trails only about 70% of the nodes are checked and some nodes are very often checked about every second time-step. With the trails systems, all nodes are regularly checked and too often checks are successfully prevented where the required computational power is reduced by 60%.

Unfortunately, when the network is fragmented in several parts and these parts are connected by a few connections, the trails of entities do not work properly. The reason for this is that many entities must use these connections, they receive a high value, and the probability to move over these connections is reduced. Fortunately, this problem can be solved in using other functions in these nodes: one solution is to use a decreasing function with a significant higher gradient or to not use the trails in this node, which results to a probability of $\frac{1}{\#neighbours}$ for all neighbours.

## 5 Analysis

In the analysis of the two techniques, the focus lays on the question how the selected information helps the entities and what other information might be interesting. In addition, the required resources are analysed and a balancing between these two parameters are done.

*Definition:* Local information comes from the network node and is not sent over the network. Global information describes the situation in another node and is sent over the network.

*Question:* Are the information locally or globally?

In the notification of insufficient security, the information is local and global because the entities receive the information from the neighbouring nodes directly and using forwarding from nodes that are nearby. However, using feasible parameters, the entities do not receive information about insufficient security in nodes that are far away. The reason for this is that if they receive information about all nodes with insufficient security, the required resources are equal to the centralised approaches. Using the parameters, it is possible to trade off the proportion between the required bandwidth and the distance where information are forwarded.

In the trails of entities, the information is locally. There is no exchange between the nodes about this information and the entities receive only the information of a node.

*Question:* Does information that is more global increase the performance?

In the notification of insufficient security, using the parameters it is possible to choose between required bandwidth and global information. However, when the system is so configured that the cells receive information about all nodes with insufficient security, the entities receive too many information and will travel too often through the network. Furthermore, it is uninteresting that an entity moves through the whole network only to fix a problem at a far-away node. Then it is more interesting to solve the problem in the area around the node where only localised information is required. The tests showed that a local setting with a distribution of some hops results in the best performance and limits the bandwidth greatly. In addition, the distribution area of the notification depends on the missing security value: the bigger the missing security level, the greater is the distribution area, which is a more adaptive approach.

In the trails of entities, some global information would increase the performance. Especially in the situation explained in the last section that connections with high load will not be traversed by entities, a view to other nodes would make the decision of an entity where to move more sophisticated. Then, the entities would receive a more global view about the nearby network. However, then the system must invest more bandwidth and the selection which information to store in which node is quite difficult.

*Question:* Which other information are interesting to use for these two problems?

For the protocol of notification of insufficient security, it would be interesting to know how many artificial cells are on the way towards this node. Furthermore it would be interesting to receive information about more than one node so that the system does not only adapt to the node which has the highest value of lacking security value. However, both would implicate lots of bandwidth and high storage space for saving the information. Furthermore, the workflow in the entities would be more complex.

For the protocol of trails of entities, it would be interesting to receive information about a part of a network and not only from the neighbours of this node. However, this would again increase the required bandwidth and storage space.

*Question:* Which other problems occur in the self-management of a distributed security system?

In the notification of insufficient security, the following questions are also of interest:

– Which node has too many entities. Too many entities is defined so that the resources - mainly computational power - is not sufficient with the required resources from all entities. It is highly interesting that these entities move to the nodes with insufficient security level in order to balance the system.
– Is it possible to design moving chains where several entities move from node to node and after one time step all nodes have a sufficient security level? If yes, which information is required so that all entities can initiate this moving?

For the trails of entities, the following questions are also interesting:

– How to use these trails so that not too many checks are performed on a node? If too many checks are performed simultaneous, the node has to invest too much computational power.
– Is it maybe possible to use these trails in order to find nodes that are controlled by an intrusion in absorbing cells? The problem in distributed security systems is that an attacker occupies a node and has access to the security systems. Then, the attacker can remove all incoming entities and after some time the system will lack of too few entities. The system can analyse the trail values in the nodes that are nearby the infected node and maybe there are association between an occupied node and trail values.

These questions will be the challenges for the future research in distributed self management of distributed security systems.

## 6 Applications

Currently, the information management is used in the artificial immune system SANA, which protects a network against intrusions [9,11]. Therefore, SANA facilitates lots of small artificial cells performing autonomously the tasks required for network security - the organic computing approach. Common used security components like antivirus softwares, firewalls, packet filters, and intrusion detection systems complement the system to benefit from their performance.

The artificial cell communication enables the collaboration between the different artificial cells where the infrastructure supplies the cells with additional information. The artificial lymph nodes implement a meeting and response point for a small part of the network and the central nativity and training stations generate and release continuously novel artificial cells including the newest security techniques. With the specialised artificial cells is it possible to implement nearly all approaches of network security. The in this article introduced self management organises the cells so that the system is always properly secured and there is no weak point in the network. The information management is more sophisticated implemented where the security components and artificial cells exchange continuously status information in order to adapt the overall system and internal thresholds as well as parameters - implementation of the danger theory [12] and the tunable activation threshold [13] of the immune cells. For the response when an intrusion is found, the components develop a defense strategy and perform this or quarantine the node with informing the administrator for further tasks. Maintenance - updates, extensions, and regular self-checks - are performed by the system using specialised cells where the maintenance is automated and not proper working as well as outdated components are reported to the administrator. However, the administrator is able to access and configure all nodes and components of the system through an administration GUI. Infected nodes are identified through regular checks as well as through distributed analysis of data gathered by various security components, i.e. the distributed analysis of network packets through intrusion detection systems and artificial cells [6]. Due tos the usage of common used components, the performance is at least as good as of common used security systems. Due to the dynamic artificial cells is the system hard to predict and to attack, contains no single point of failure, and adapts to the current situation in the network. In addition, a real distributed intrusion detection system is implemented in all nodes with simplified administration workflows [9].

Furthermore, the notification of low security can be used in nearly all distributed systems where the number of entities can be measured and thereafter modeled using the parameters and functions. The trails can be used in all scenarios where a type of entity should visit all nodes regularly but not too frequently. These systems are most distributed systems as the multi-agent systems, the massively distributed systems, complex adaptive systems, and the artificial immune systems.

## 7 Conclusion

This article discusses two information protocols for distributed systems and analyses the performance. The two protocols increase the performance of a distributed system enormously because the entities receive information where their tasks are required. In addition, the required bandwidth is limited because a centralised system is not used. The next steps are to introduce more information protocols and evaluate theoretically how the performance would increase when more bandwidth is invested. Furthermore, the enhancing of the distributed security systems that base on the artificial immune system as archetype is a next step.

## Acknowledgments

The PhD-Project SANA - Security Analysis in iNternet trAffic - is financially supported by the University of Luxembourg. I like to thank Christoph Schommer for helpful discussions and the Ministre Luxembourgeois de l'education et de la recherche for additional financial support.